\documentclass[10pt,a4paper]{article}
\usepackage[T1]{fontenc}
\usepackage{amsmath,mathtools,amssymb}
\usepackage{graphicx,relsize}
\usepackage{newtxtext,newtxmath} % 
\usepackage{multirow,lineno}
\usepackage{xcolor}
\usepackage{titlesec,etoolbox}
\usepackage{siunitx,booktabs,tabularx,ragged2e,threeparttable}
\usepackage[all]{xy}
\usepackage{microtype}
\usepackage{setspace}
\usepackage{url}
\usepackage[authoryear, round]{natbib}
\setcitestyle{aysep={,}, yysep={,}, notesep={, }}

\usepackage{hyperref}
\hypersetup{
	colorlinks=true,
	linkcolor=red,
	citecolor=red,
	urlcolor=red,
	pdfencoding=auto,  % 
	psdextra,          % 
	breaklinks=true,   % 
	pdfborder={0 0 0}, % 
}

\usepackage[left=2.cm,right=2.cm,top=2.cm,bottom=2.cm]{geometry}

\selectfont

\titleformat{\section}
{\fontsize{9}{9}\selectfont\bfseries\raggedright}
{\thesection.}
{0.3em}
{}
\titlespacing*{\section}{0pt}{9pt}{6pt}

\titleformat{\subsection}
{\fontsize{9}{9}\selectfont\bfseries\raggedright}
{\thesubsection.}
{0.3em}
{}
\titlespacing*{\subsection}{0pt}{9pt}{6pt}

\titleformat{\subsubsection}
{\fontsize{9}{9}\selectfont\bfseries\itshape\raggedright}
{\thesubsubsection.}
{0.3em}
{}
\titlespacing*{\subsubsection}{0pt}{6pt}{3pt}

\title{\textcolor{black}{Multi-factor modeling of chlorophyll-a in South China’s subtropical reservoirs using long-term monitoring data for quantitative analysis}}
\author{
	Haizhao Guan\textsuperscript{1}, 
	Yiyuan Niu\textsuperscript{2}, 
	Chuanjin Zu\textsuperscript{3}, 	
	Ju Kang\textsuperscript{4,*}
}
\date{}
\newcommand{\affil}[2]{%
	\textsuperscript{#1}#2%
}
\makeatletter
\renewcommand{\maketitle}{
	\begin{center}
		{\LARGE\bfseries \@title\par}
		\vspace{1em}  % 
		{\normalsize \@author\par}
		\vspace{-0.3em} % 
		\setlength{\parskip}{1pt}
		\@date
	\end{center}
}
\makeatother
\begin{document}	
	\maketitle
	\begin{center}
		\affil{1}{Guangzhou Bureau of Hydrology, Guangdong Provincial Bureau of Hydrology, Guangzhou 510150, China}\\
		\affil{2}{School of Physics, Sun Yat-sen University, Guangzhou 510275, China}\\
		\affil{3}{Ocean College, Jiangsu University of Science and Technology, Zhenjiang 212100, China}\\		
		\affil{4}{School of Ecology, Sun Yat-sen University, Shenzhen 518107, China}\\
		\affil{*}{Corresponding author: \href{mailto:kangj29@mail.sysu.edu.cn}{kangj29@mail.sysu.edu.cn}}
	\end{center}	
	\begin{abstract}
		\textcolor{black}{Eutrophication and harmful algal blooms, driven by complex interactions among nutrients and climate, threaten freshwater ecosystems globally, particularly in densely populated Asian regions where rapid urbanization and agricultural intensification exacerbate nutrient pollution.} \textcolor{black}{Understanding the non-linear interactions among water temperature, nutrient levels, and chlorophyll-a (Chl-a) dynamics is crucial for addressing eutrophication in freshwater ecosystems. Many existing studies, however, tend to oversimplify these relationships and lack validation with long-term field data.} Here, we conducted multi-year field monitoring (2020-2024) of key environmental factors, including total nitrogen (TN), total phosphorus (TP), water temperature, and Chl-a, across three reservoirs in Guangdong Province, China: Tiantangshan (S1), Baisha River (S2), and Meizhou (S3). \textcolor{black}{Chl-a concentrations showed significant spatiotemporal variability, ranging from 1.2 to 11.8 $\mu\text{g}/\text{L}$, with a general increasing trend indicative of progressing eutrophication.} Strong positive correlations were found between Chl-a and TN, TP, and temperature. \textcolor{black}{Numerical analysis of the long-term data revealed TN as a more influential driver than TP for Chl-a proliferation in these systems, with Chl-a increasing by an average of 4.2 $\mu\text{g}/\text{L}$ per unit increase in TN (mg/L), compared to 2.8 $\mu\text{g}/\text{L}$ per unit increase in TP (mg/L).} Based on the collected data, we developed and calibrated a dynamic multi-factor hydro-ecological model. \textcolor{black}{The model accurately reproduced the observed Chl-a patterns (R\textsuperscript{2} > 0.85), identifying synergistic effects between temperature and nutrients, particularly a 15\% enhancement in Chl-a growth rate when temperature exceeded 25 \textcelsius~concurrent with high TN.} \textcolor{black}{The model offers a robust theoretical basis for predicting Chl-a dynamics and supports science-informed management strategies, such as targeted nitrogen load reduction, to mitigate eutrophication in subtropical reservoirs. This study underscores the necessity of multi-factor, process-based approaches for sustainable water resource management under changing climatic and anthropogenic pressures.}
		\vspace{0.3cm}	
	\end{abstract}	
	\section{Introduction}
	Aquatic ecosystems play a fundamental role in supporting life on Earth,
	with their structure and functioning shaped by complex interactions among key environmental factors, particularly water temperature, nutrient availability (e.g., total nitrogen (TN) and total phosphorus (TP)), and algal biomass, typically represented by chlorophyll-a (Chl-a). In recent decades, global climate change and intensified human activity have profoundly altered these ecosystems, driving more frequent extreme hydrological events, accelerating eutrophication, and contributing to widespread biodiversity loss~\citep{PhilFong2025,PriyaAK2023,Donat-PHader2019,SusanneMenden-Deuer2025,ElisaSoana2024,ZihanZhao2025,RicardoHirata2025,HulyaBoyacioglu2024}. \textcolor{black}{In Asia, rapid urbanization and agricultural intensification have exacerbated nutrient pollution in freshwater bodies. Studies from South Asia, including Pakistan and Bangladesh, document severe eutrophication pressures in river basins and lakes, highlighting a regional crisis that demands locally calibrated understanding and solutions}~\citep{AyazUlHaq2023, SaidMuhammad2023, CemTokatli2025, CemTokatli2023, SaidMuhammad2024}. \textcolor{black}{Specifically, in China's subtropical regions, reservoirs are experiencing increasing frequency and intensity of algal blooms due to combined pressures of excessive nutrient loading from watersheds and rising water temperatures associated with climate change. Effectively managing these systems requires a quantitative understanding of how these multiple drivers interact to control phytoplankton biomass.}
	
	\textcolor{black}{A critical challenge in predicting algal dynamics lies in the non-linear and often synergistic interactions among multiple stressors, such as warming and nutrient enrichment. Climate change can exacerbate the impacts of nutrient pollution, altering phytoplankton community structure and bloom thresholds in ways that single-factor models fail to capture}~\citep{Mpakairi2024, Latwal2024}. These environmental shifts underscore the urgent need to unravel the interdependencies among multiple stressors and to quantify their ecological impacts through robust, mechanistic models
	~\citep{Liang2025, PengQi2025, Huang2022}.
	
	Although multi-factor modeling in aquatic ecology has advanced in recent years,
	many models remain limited in \textcolor{black}{predictive and} explanatory power.
	Early studies typically focused on linear, single-factor relationships,
	such as nutrient concentrations and Chl-a growth. However, the emerging field of complexity science has prompted a shift toward integrative approaches that capture interactions among multiple drivers. For instance, Liu et al.~\citep{Liu2021} employed a coupled hydrodynamic-ecological model to quantify the interactive effects of TN, TP, total suspended solids (TSS), and light availability on phytoplankton competition in Chagan Lake, identifying TSS as a primary driver of cyanobacterial succession. Similarly, Qian et al.~\citep{JingQian2024} applied deep learning to model algal bloom dynamics in Taihu Lake, revealing synergistic effects between thermal stratification and nutrient input. Zhang et. al.~\citep{HanxiaoZhang2024} explored spatial patterns of TP and Chl-a in reservoirs, highlighting the influence of geographic and meteorological factors such as latitude, slope, and temperature on Chl-a variability. \textcolor{black}{Remote sensing studies in subtropical reservoirs, such as the Nandoni reservoir in South Africa, have further corroborated the utility of high-resolution monitoring in capturing the spatial heterogeneity of Chl-a, often revealing higher concentrations near inflows and reservoir edges, which aligns with the need for spatially explicit understanding of eutrophication drivers}~\citep{Mpakairi2024}. \textcolor{black}{Systematic reviews emphasize the growing effort to develop integrated models that couple watershed processes with in-lake dynamics to support management}~\citep{Shi2024, Buta2023}. \textcolor{black}{In South Asia, research in the Ganges-Brahmaputra and Indus River basins has also highlighted the critical role of anthropogenic nutrient loads and climate variability in driving aquatic ecosystem changes}~\citep{SaidMuhammad2024,CemTokatli2023}.
	
	While these studies contribute valuable data and demonstrate the potential of advanced
	modeling techniques, they often rely heavily on statistical or machine learning approaches, offering limited insight into the underlying ecological mechanisms
	~\citep{Liu2021,JingQian2024,HanxiaoZhang2024}. \textcolor{black}{A key limitation of many existing models lies in their calibration with short-term datasets and the consequent lack of rigorous validation against long-term observational records; this deficiency critically undermines their ability to accurately simulate and predict system responses under realistic conditions of interannual climatic variability and episodic disturbance events—such as the extreme rainfall prevalent in subtropical monsoonal regions—which can drastically alter nutrient fluxes and phytoplankton dynamics} \citep{Buta2023}. Moreover, traditional models such as the Water Erosion Prediction Project (WEPP), although effective in simulating physical processes like soil erosion, fall short in capturing hydro-ecological interactions \citep{Flanagan2012,ZhuoxinChen2025}.
	Some progress \textcolor{black}{has} been made by integrating robust optimization and scenario analysis into water resource management, for example, Xu et al.~\citep{BinXu2022} incorporated multi-objective optimization and copula-based uncertainty quantification into cascade reservoir planning, but the ecological models \textcolor{black}{remain} relatively simplified. In contrast, emerging dynamic multi-factor models offer promising tools to explore non-linear feedbacks and temporal continuity among interacting ecological variables, including Chl-a growth, nutrient cycling, and temperature dynamics. Despite their potential, these models are still in early stages of development and are seldom applied to real-world aquatic systems.
	
	\textcolor{black}{To address these issues, this study aims to: (1) quantify the spatiotemporal dynamics and key drivers of Chl-a in three subtropical reservoirs using a comprehensive five-year (2020-2024) monitoring dataset; (2) develop a process-based, dynamic multi-factor model that mechanistically integrates the effects of water temperature, TN, and TP on Chl-a; and (3) calibrate and validate the model against the long-term data to assess its predictive capability and provide a tool for exploring management scenarios.} We propose a dynamic, multi-factor modeling approach to investigate
	the interactions among water temperature, TN, TP, and Chl-a in three freshwater reservoirs in Guangdong, China. By leveraging long-term monitoring data and integrating ecological mechanisms into the modeling framework, this work aims to deepen our understanding of eutrophication processes and provide a scientific basis for informed aquatic ecosystem management.	
	\begin{figure}[ht!]   
		\centering
		\includegraphics[width=15cm]{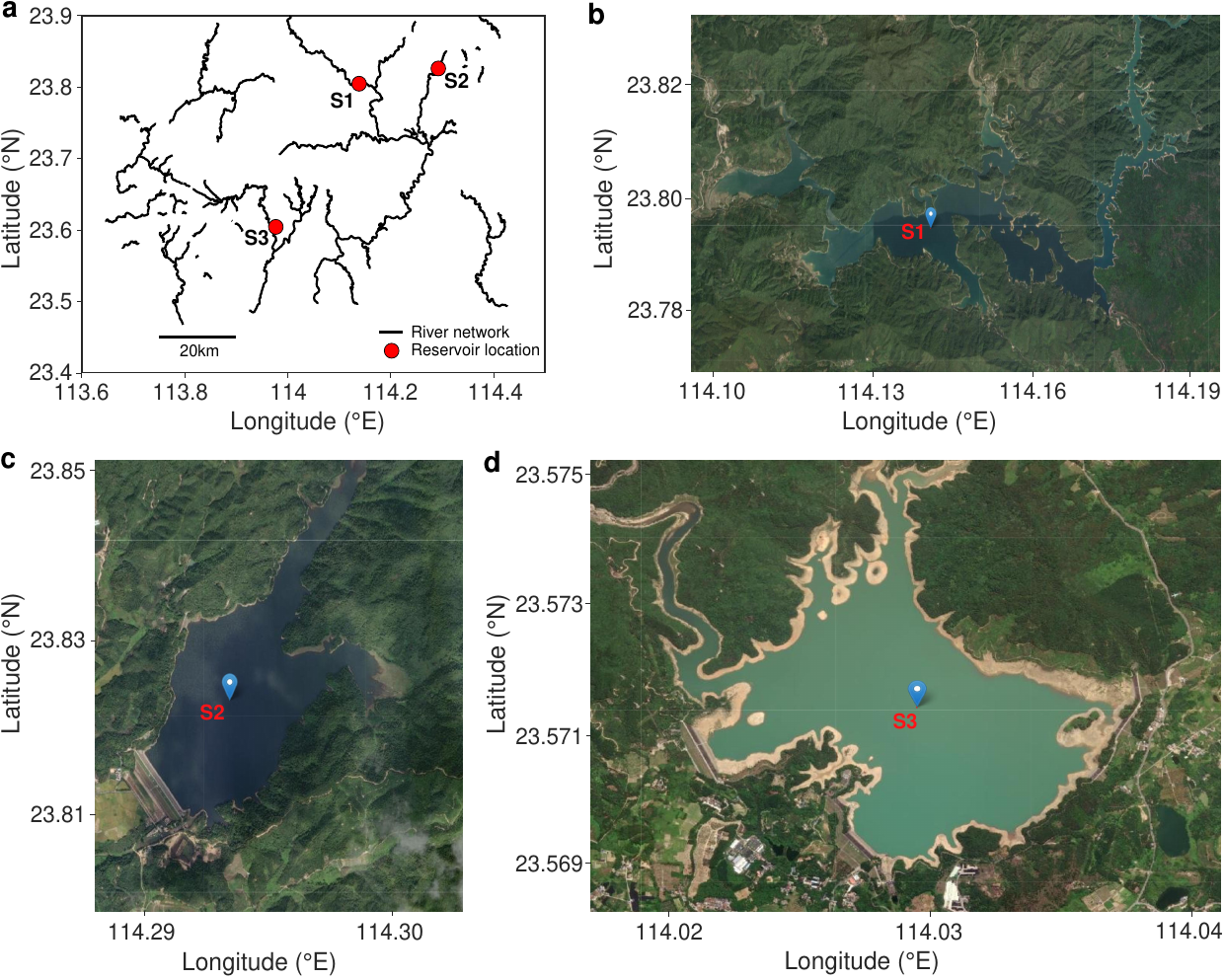}
		\caption{\label{fig1} 
			Locations of the three studied reservoirs \textcolor{black}{(S1: Tiantangshan; S2: Baisha River; S3: Meizhou)} in Guangdong Province, China. The inset \textcolor{black}{map shows their geographical context} within the Pearl River Basin.}   
	\end{figure}	
	\section{Materials and methods}
	\subsection{Research sites and sample collection}
	As illustrated in Fig.~\ref{fig1}, Tiantangshan Reservoir (S1: 114.17 \textdegree E, 23.79 \textdegree N), Baisha River Reservoir (S2: 114.29 \textdegree E, 23.82 \textdegree N), and Meizhou Reservoir (S3: 114.03 \textdegree E, 22.57 \textdegree N) are located within the upper reaches of the Zengjiang River, a major tributary of the Dongjiang River Basin in Guangdong Province, China. This region experiences a typical subtropical monsoon climate. Between January 2020 and December 2024, water quality samples were collected. \textcolor{black}{Sampling frequency differed: monthly at S1, and quarterly at S2 and S3. This difference is acknowledged as a limitation for direct temporal comparability, and analyses were designed to focus on seasonal and interannual patterns within each site and on comparative spatial assessments.} A total of 100 samples were collected across the three sites (see Table~\ref{tab1}).
	
	\begin{table}[htbp]
		\centering
		\caption{Geographic coordinates and sampling details for the three reservoirs.}
		\label{tab1}
		\setlength{\tabcolsep}{4pt}
		\begin{tabularx}{\textwidth}{
				>{\hsize=1.3\hsize\centering\arraybackslash}X
				>{\hsize=0.9\hsize\centering\arraybackslash}X
				>{\hsize=0.9\hsize\centering\arraybackslash}X
				>{\hsize=0.9\hsize\centering\arraybackslash}X
				>{\hsize=1.0\hsize\centering\arraybackslash}X
			}
			\toprule
			\textbf{Reservoir Name} & \textbf{Code} & \textbf{Longitude (\textdegree E)} & \textbf{Latitude (\textdegree N)} & \textbf{Main Tributary} \\
			\midrule
			Tiantangshan Reservoir & S1 & 114.17 & 23.79 & Xilin River \\
			Baisha River Reservoir     & S2 & 114.29 & 23.82 & Baisha River \\
			Meizhou Reservoir      & S3 & 114.03 & 22.57 & Yonghan River \\
			\bottomrule
		\end{tabularx}
	\end{table}	
	\subsection{Determination of physical and chemical properties}
	\textcolor{black}{In-situ measurements of water temperature, pH, dissolved oxygen (DO), and electrical conductivity were performed using a multi-parameter sonde (YSI EXO2, USA).} TN was analyzed using the continuous flow spectrophotometric method (HJ667-2013). TP was determined via the ammonium molybdate spectrophotometric method (HJ670-2013). Nitrate-nitrogen (\textcolor{black}{NO\textsubscript{3}-N}) was measured by ion chromatography (SL86-1994). Chl-a concentrations were determined by acetone extraction and spectrophotometry (SL88-2012). 
	
	\subsection{Data analysis}
	Principal component analysis (PCA) was employed to elucidate the major patterns and key drivers within the dataset. The analysis included the following water quality parameters: TN, TP, Chl-a, water temperature, pH, and DO. Prior to PCA, all variables were standardized to zero mean and unit variance. Principal components (PCs) with eigenvalues greater than 1 were retained, with the first two PCs cumulatively explaining over 70\% of the total variance in the data. All statistical computations and visualizations were performed using MATLAB R2019a.	
	\subsection{Model and stability analysis}	
	\textcolor{black}{The preceding univariate and multivariate analyses identified water temperature, TN, and TP as the dominant co-varying drivers of Chl-a dynamics in the studied reservoirs. To mechanistically integrate these interactions and quantify their nonlinear feedbacks, we developed a dynamic hydro-ecological model. This model couples nutrient cycling and phytoplankton growth, providing a process-based framework to simulate the temporal co-evolution of TN, TP, and Chl-a, and to assess ecosystem stability under varying environmental conditions.}
	
	\textcolor{black}{The model is formulated as a system of three ordinary differential equations representing the rates of change for TN concentration $N$ (mg/L), TP concentration $P$ (mg/L), and Chl-a concentration $C$ ($\mu\text{g}/\text{L}$):}
	\begin{equation}
		\begin{cases}
			\begin{aligned}
				\frac{dN}{dt} &= 
				\underbrace{Q_{N}(N_{in}-N)}_{\mathclap{\text{Net inflow}}} 
				- \overbrace{k_{1}(T)N}^{\mathclap{\text{Sedimentation loss}}} 
				+ \underbrace{S_{N}}_{\mathclap{\text{Sediment release}}} 
				+ \overbrace{\gamma_{1}C}^{\mathclap{\text{~~~~~~~~~~~~Decomposition release}}} 
				- \underbrace{\alpha_{1}g_{1}(N)C}_{\mathclap{\text{Consumption loss}}} 
				\\[2ex]  % 增加行间距		
				\frac{dP}{dt} &= 
				\underbrace{Q_{P}(P_{in}-P)}_{\mathclap{\text{Net inflow}}} 
				- \overbrace{k_{2}(T)P}^{\mathclap{\text{Sedimentation loss}}} 
				+ \underbrace{S_{P}}_{\mathclap{\text{Sediment release}}} 
				+ \overbrace{\gamma_{2}C}^{\mathclap{\text{~~~~~~~~~~~~Decomposition release}}} 
				- \underbrace{\alpha_{2}g_{2}(P)C}_{\mathclap{\text{Consumption loss}}} 
				\\[2ex]		
				\frac{dC}{dt} &= 
				\underbrace{wg_{1}(N)g_{2}(P)}_{\mathclap{\text{Absorption and transformation}}} 
				\overbrace{C\left(1-\frac{C}{K}\right)}^{\mathclap{\text{Logistic growth}}} 
				- \underbrace{dC}_{\mathclap{\text{~~~~~~Natural mortality}}}
			\end{aligned}
		\end{cases}
		\label{model}
	\end{equation}
	The key parameters and their definitions used in the Eq.~(\ref{model}) are summarized in Table~~\ref{tab2}. \textcolor{black}{The functions $g_i(X) = \dfrac{X}{K_X + X}$ represent Michaelis-Menten nutrient uptake kinetics, where $K_X$ is the half-saturation constant. The temperature dependence of sedimentation is modeled as $k_i(T) = k_{i0} \theta^{(T - 20)}$, an Arrhenius-type function. All state variables and parameters are constrained to be non-negative, confining the system dynamics to the biologically relevant state space $\mathbb{R}_{+}^{3}=\left\{ \left( N,P,C \right)\in {{\mathbb{R}}^{3}}:N\ge 0,P\ge 0,C\ge 0 \right\}$}.
	\begin{table}[htbp]
		\centering
		\caption{Definitions of parameters used in the aquatic ecosystem model.}
		\label{tab2}
		\begin{threeparttable}
			\begin{tabularx}{\textwidth}{lXc}  % 使用X列自动调整宽度
				\toprule
				Symbols & Illustrations  \\
				\midrule
				$Q_{N}, Q_{P}$ & Net inflow rates of \textcolor{black}{TN} and \textcolor{black}{TP} \\
				$N_{in}, P_{in}$ & External inputs of \textcolor{black}{TN} and \textcolor{black}{TP} \\
				$k_1(T), k_2(T)$ & Temperature-dependent sedimentation rate constants for nitrogen and phosphorus \\
				%$k_i(T) = k_{i0} \theta^{(T - 20)}$ & Arrhenius-type function for temperature dependence of $k_i$ \\
				$k_{i0}$ & Base sedimentation rate constant \\
				$\theta$ & Empirical temperature coefficient (typically 1.02 to 1.06) \\
				$T$ & Water temperature (\si{\degreeCelsius}) \\
				$S_N, S_P$ & Release rates of nitrogen and phosphorus from sediments \\
				$\alpha_1, \alpha_2$ & Nutrient consumption rates by \textcolor{black}{Chl-a}\\	
				$w$ & Conversion coefficient from nutrient consumption to \textcolor{black}{Chl-a} biomass \\
				$K$ & Maximum carrying capacity of \textcolor{black}{Chl-a} \\
				$d$ & Natural degradation rate of \textcolor{black}{Chl-a} \\
				%$g_i(X) = \dfrac{X}{K_X + X}$ & Michaelis-Menten uptake function; $K_X$ is the half-saturation constant \\
				\bottomrule
			\end{tabularx}
			\begin{tablenotes}
				\item Note that in the table, $i=1,2$.
			\end{tablenotes}
		\end{threeparttable}
	\end{table}
	
	We set $dN/dt = 0$, $dP/dt = 0$, and $dC/dt = 0$ to find the unique coexistent equilibrium 
	$E(N^{*},P^{*},C^{*})$ of system~(\ref{model}). This equilibrium point $E$ satisfies the following 
	algebraic equations:
	\begin{equation}
		\begin{cases}	
			\vphantom{}Q_{N}(N_{in}-N^{*}) -k_{10}\theta^{(T-20)}N^{*} +S_{N} + \gamma_{1}C^{*}-\alpha_{1}\dfrac{N^{*}}{K_{N}+N^{*}}C^{*} = 0 \\[1.5ex]  % 调整 [1.5ex] 数值控制间距		
			\vphantom{} Q_{P}(P_{in}-P^{*}) -k_{20}\theta^{(T-20)}P^{*} +S_{P}
			+ \gamma_{2}C^{*}-\alpha_{2}\dfrac{P^{*}}{K_{P}+P^{*}}C^{*}  = 0  \\[1.5ex]  % 例如 [2ex] 间距更大
			\vphantom{}  w\dfrac{N^{*}}{K_{N}+N^{*}}\dfrac{P^{*}}{K_{P}+P^{*}}\left( 1-\dfrac{C^{*}}{K}\right) -d = 0
		\end{cases}
		\label{stable}
	\end{equation}
	Next, we analyse the local asymptotic stability of system~(\ref{model}) at the coexistent equilibrium $E$
	using the Jacobian matrix and the Routh-Hurwitz criterion. 
	The Jacobian matrix of system~(\ref{model}) evaluated at $E$ is given by
	\begin{equation}
		J({{E}})={{D}_{{{X}_{j}}}}{{f}_{i}}({{x}_{j}})\left| _{{{E}}} \right.={{\left[ \frac{\partial {{f}_{i}}}{\partial {{X}_{j}}} \right]}_{{{E}}}},(i,j=1,2,3)
		\label{Jacobian_Eq}
	\end{equation}
	Then the characteristic equation corresponding to the Jacobian matrix $J(E)$ is
	\begin{equation}
		{{\lambda }^{3}}+{{a}_{2}}{{\lambda }^{2}}+{{a}_{1}}\lambda +{{a}_{0}}=0
		\label{characteristic_Eq}
	\end{equation}
	Where $e_{ij}$ denotes the element in the $i$-th row and $j$-th column of the Jacobian matrix~(\ref{Jacobian_Eq}),
	and the coefficients $a_2$, $a_1$, and $a_0$ are given by:
	\begin{equation}
		\begin{cases}
			{{a}_{2}}={{e}_{\text{11}}}+{{e}_{\text{22 }}}+{{e}_{\text{33}}}\\
			{{a}_{1}}={{e}_{\text{12}}}{{e}_{\text{21}}}+{{e}_{\text{13}}}{{e}_{\text{31}}}+{{e}_{\text{23}}}{{e}_{\text{32}}}-{{e}_{\text{22}}}{{e}_{\text{33}}}-{{e}_{\text{11}}}\left( {{e}_{\text{22}}}+{{e}_{\text{33}}} \right)\text{ }\\
			{{a}_{0}}={{e}_{\text{12}}}{{e}_{\text{23}}}{{e}_{\text{31}}}+{{e}_{\text{13}}}{{e}_{\text{21}}}{{e}_{\text{32}}}-{{e}_{\text{11}}}{{e}_{\text{23}}}{{e}_{\text{32}}}-{{e}_{\text{12}}}{{e}_{\text{21}}}{{e}_{\text{33}}}+{{e}_{\text{11}}}{{e}_{\text{22}}}{{e}_{\text{33}}}-{{e}_{\text{13}}}{{e}_{\text{22}}}{{e}_{\text{31}}}
		\end{cases}
		\label{element}
	\end{equation}
	
\begin{figure}[ht!]
		\centering
		\includegraphics[width=15cm]{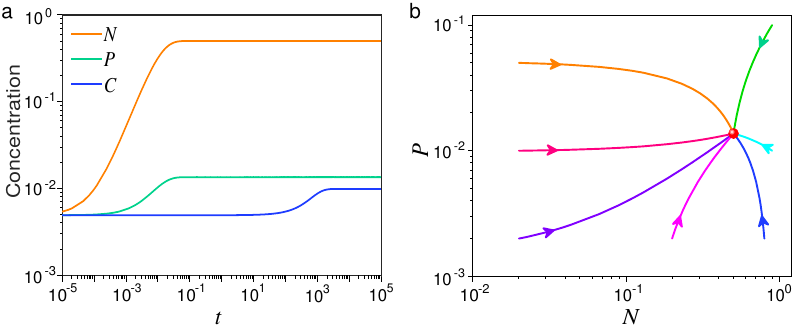}
		\caption{\label{fig:stability}
			\textcolor{black}{Numerical demonstration of local asymptotic stability for the dynamic model (Eq.~\ref{model}). (a) Time series showing the convergence of TN (\( N \)), TP (\( P \)), and Chl-a (\( C \)) concentrations to the stable equilibrium \( E \). (b) Phase portrait in the \( N \)-\( P \) plane, with arrows indicating the direction of system trajectories towards the stable focus at \( E \). The simulation parameters are set as follows: $k_{10} = 0.10$, $ k_{20} = 0.6$, $\gamma_{1} = 0.3$, $ \gamma_{2} = 0.1$, $\alpha_{1} = 0.3$, $\alpha_{2} = 0.1$, $Q_{N} = 0.01$, $Q_{P} = 0.01$, $N_{in} = 0.5$, $P_{in} = 0.02$, $S_{N} = 0.5$, $ S_{P} = 0.1$, $K_{N} = 0.5$, $K_{P} = 0.5$, $T = 25$, $d = 0.002 $, $\theta = 1.04$, $K = 0.02$, $w = 0.3$.}}
	\end{figure}

	According to the Routh-Hurwitz criterion, the system~(\ref{model}) is locally asymptotically stable 
	at the coexistent equilibrium $E$ if the following conditions hold simultaneously: $a_1>0, \quad a_2>0, \quad a_0>0 \quad \text{and} \quad a_1 a_2>a_0$. If any of these conditions fails, the equilibrium $E$ is unstable. \textcolor{black}{Using the parameter set (as specified in the caption of Fig.~\ref{fig:stability}), the system converges to a stable fixed point $E$(see Fig.~\ref{fig:stability}). At this equilibrium, all eigenvalues of the Jacobian matrix \( \mathbf{J}(E) \) are negative real parts, satisfying the condition for local asymptotic stability. This convergence is visualized in Fig.~\ref{fig:stability}, where trajectories from different initial conditions approach the stable fixed point in both time series and phase space.}
	\section{Results}
	\subsection{\textcolor{black}{Spatiotemporal dynamics of water quality parameters}}
	\textcolor{black}{Between 2020 and 2024, the three monitored reservoirs exhibited distinct yet interconnected trends in key water quality parameters. Water temperature followed predictable seasonal cycles, while concentrations of TN, TP, and Chl-a showed varying degrees of interannual fluctuation and spatial difference, indicating a gradual shift in trophic state.}
	
\begin{figure}[ht!]
		\centering
		\includegraphics[width=17cm]{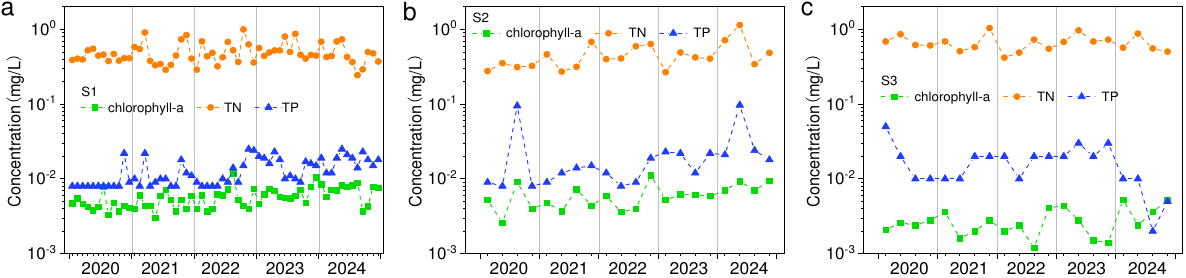}
		\caption{\label{Ye-aTNTP}
			\textcolor{black}{Interannual variations (2020-2024) of total nitrogen (TN), total phosphorus (TP), and Chl-a concentrations in Tiantangshan (S1), Baisha River (S2), and Meizhou (S3) reservoirs. Note the distinct seasonal temperature cycle, the high variability in TN, and the overall increasing trend in Chl-a.}}
	\end{figure}

\begin{figure}[ht!]
		\centering
		\includegraphics[width=15cm]{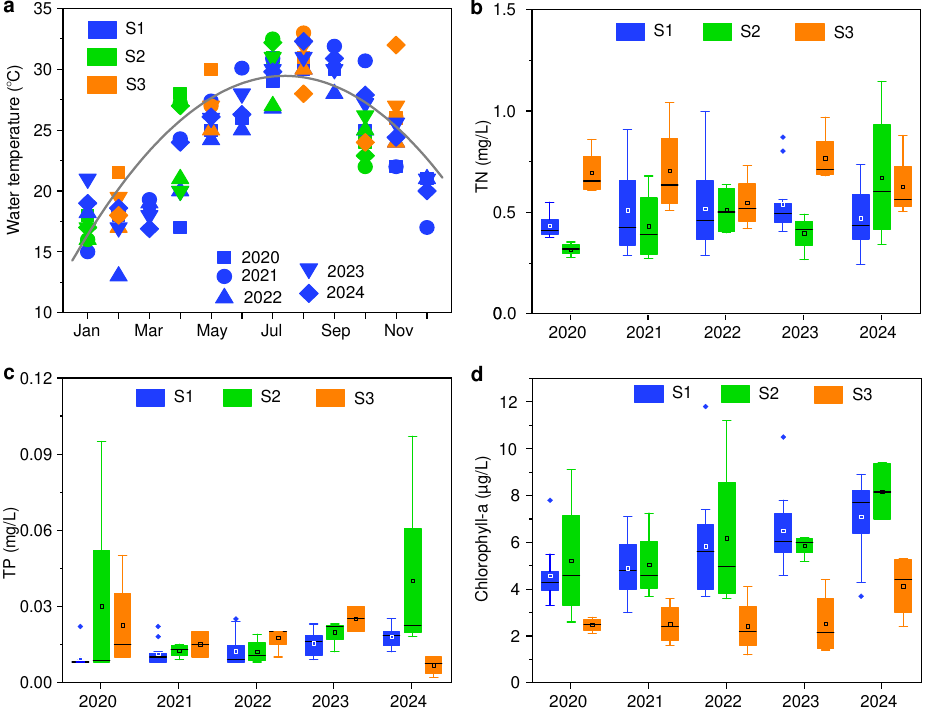}
		\caption{\label{xiangxiantu}
			\textcolor{black}{Detailed temporal trends of (a) water temperature, (b) TN, (c) TP, and (d) Chl-a from 2020 to 2024. The plots highlight the seasonal consistency of temperature, episodic nutrient peaks (particularly in S2), and the synchronous rise in Chl-a and TN in S1 and S2 during 2022. }}
	\end{figure}

	\textcolor{black}{Water temperature exhibited a consistent seasonal pattern across all reservoirs, peaking during the summer months (July–September) (Figs.~\ref{Ye-aTNTP}a,~\ref{xiangxiantu}a). Meizhou Reservoir (S3) maintained the highest annual average temperatures throughout the study period (ranging from 24.0 to 27.1 \textcelsius), followed by Baisha River Reservoir (S2: 22.3-25.3 \textcelsius) and Tiantangshan Reservoir (S1: 22.9-24.9 \textcelsius). This spatial gradient suggests potential influences from local microclimate, reservoir morphology, or inflow characteristics.}
	
	\textcolor{black}{TN concentrations displayed the most pronounced interannual and spatial variability (Fig.~\ref{Ye-aTNTP}b-d). S3 consistently had the highest TN levels (e.g., 0.68-0.97 mg/L in 2023), while S1 and S2 generally showed lower concentrations with greater fluctuations. Notably, sharp increases were observed in S2 in 2024 (peak: 1.15 mg/L) and in S1 during 2021-2022, coinciding with periods of elevated Chl-a. In contrast, TP concentrations were more stable, particularly in S1 and S3 (Fig. \ref{xiangxiantu}c). S2, however, experienced marked TP spikes in 2020 and 2024 (peaks up to 0.097 mg/L), indicating potential episodic external loading or internal sediment release events at this site.}
	
	\textcolor{black}{Chl-a dynamics revealed critical insights into phytoplankton biomass and eutrophication risk (Figs.~\ref{Ye-aTNTP},~\ref{xiangxiantu}d). A gradual but consistent upward trend in Chl-a concentrations was observed across all three reservoirs from 2020 to 2024. S1 and S2 exhibited higher and more variable Chl-a levels, with distinct peaks in 2022 (S1: 11.8 $\mu\text{g}/\text{L}$; S2: 11.2 $\mu\text{g}/\text{L}$). In contrast, S3 maintained consistently low Chl-a concentrations (<5.3 $\mu\text{g}/\text{L}$), despite its higher TN levels. This dissociation suggests that factors beyond TN (such as light limitation, grazing pressure, or phosphorus availability) may strongly regulate algal growth in S3. The synchronous peaks of TN and Chl-a in S1 and S2 in specific years (e.g., 2022) point to nitrogen-driven productivity pulses in these systems.}
	
	\textcolor{black}{The co-variation of TN and Chl-a in S1 and S2 highlights nitrogen as a potential key driver of phytoplankton biomass in these reservoirs. The significant TP spikes in S2, especially in 2024, did not coincide with proportional Chl-a increases, suggesting a possible shift from phosphorus limitation or the influence of other mitigating factors. Overall, the data indicate an increasing trajectory of eutrophication pressure across the studied reservoirs, with S1 and S2 showing higher vulnerability to algal biomass accumulation linked to nitrogen dynamics.}
	
	\subsection{\textcolor{black}{Principal component analysis (PCA) of environmental factors in three reservoirs (2020-2024)}}
	\textcolor{black}{PCA was applied to elucidate the primary environmental gradients and their temporal trends across the three reservoirs. The analysis integrated six key water quality parameters: TN, TP, Chl-a, water temperature, pH, and DO.}
	
\begin{figure}[ht!]
		\centering
		\includegraphics[width=15cm]{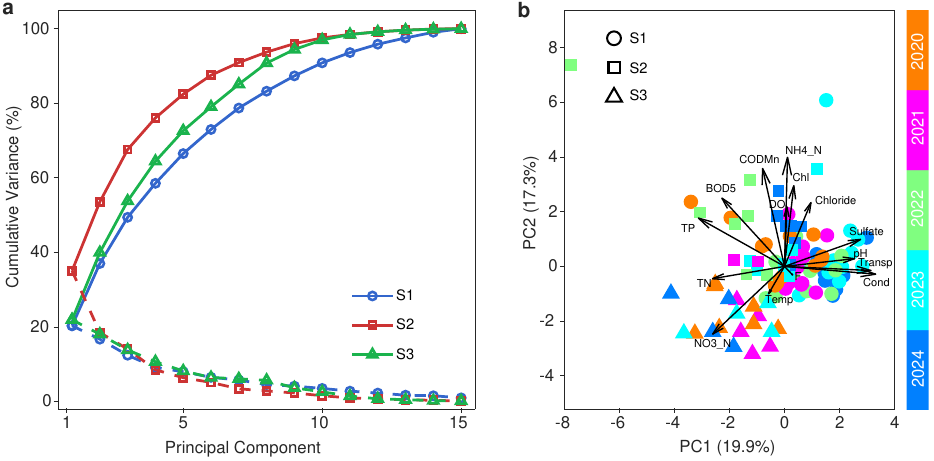}
		\caption{\label{PCA}
			\textcolor{black}{PCA of environmental variables in Tiantangshan (S1), Baisha River (S2), and Meizhou (S3) reservoirs (2020–2024). (a) Scree plot showing the variance explained by the first five principal components. The first two PCs (PC1 and PC2) explain >70\% of the total variance. (b) Biplot of sample scores (points) and variable loadings (arrows) in the PC1-PC2 space. The trajectory of points from 2020 to 2024 shows a shift along the PC1 axis, correlated with increasing Chl-a. Vectors for TN and TP align with PC2.}
		}
	\end{figure}

	\textcolor{black}{The scree plot (Fig.~\ref{PCA}a) shows that the first two principal components (PCs) collectively explained over 70\% of the total variance within the dataset, with PC1 contributing the majority. This indicates that these two components effectively capture the predominant patterns of co-variation among the measured variables, providing a robust basis for dimensional reduction and interpretation.}
	
	\textcolor{black}{The biplot of PC1 versus PC2 (Fig.~\ref{PCA}b) reveals distinct environmental trajectories and spatial differentiation. Temporally, samples exhibited a consistent rightward shift along the PC1 axis from 2020 to 2024. This trajectory is strongly aligned with the positive loading of Chl-a on PC1, visually confirming the long-term increasing trend in phytoplankton biomass identified in the univariate analysis. Spatially, samples from the three reservoirs formed distinguishable clusters. Samples from S3 clustered tightly in the lower quadrant, reflecting its relatively stable and distinct water quality character. In contrast, samples from S1 and S2 showed greater dispersion along both PC axes, indicating higher temporal variability in their environmental conditions.}
	
	\textcolor{black}{The variable loadings provide insight into the drivers behind these patterns. PC1 is predominantly associated with Chl-a and water temperature, representing a gradient of biological activity and thermal energy. PC2 is strongly positively loaded with TN and TP, representing a gradient of nutrient enrichment. 
		The orthogonal positions of the nutrient vectors (TN, TP) relative to the Chl-a vector on the biplot suggest that while nutrient concentrations are a key source of variation among samples, the coupling between nutrient levels and instantaneous algal biomass (Chl-a) is complex and potentially mediated by other factors captured along different axes.}
	
	\subsection{Quantitative analysis of chlorophyll-a concentration in relation to water temperature}
	\textcolor{black}{Water temperature is a fundamental regulator of phytoplankton physiology and growth. Building upon established knowledge of its complex effects on algal metabolism and community dynamics}~\citep{Gobler2020, EijiMaeda2019, PaerlHW2009}. \textcolor{black}{We quantitatively assessed this relationship using our dynamic model (Eq.~\ref{model}). The model explicitly incorporates temperature dependence by parameterizing the growth rate coefficient $w$ as a function of water temperature $T$, i.e., $w = w(T)$. }
	
\begin{figure}[ht!]
		\centering
		\includegraphics[width=17cm]{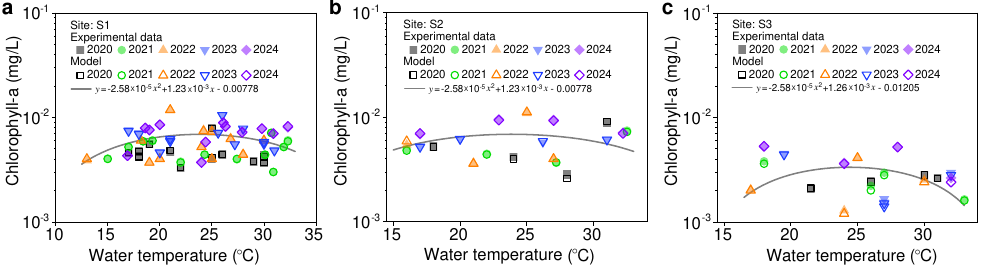}
		\caption{\label{S1-S3}
			\textcolor{black}{Model-simulated relationship between water temperature and Chl-a concentration for Tiantangshan (S1), Baisha River (S2), and Meizhou (S3) reservoirs. The curves depict a characteristic unimodal response, with optimal growth occurring within a broad temperature window.}
		}
	\end{figure}

	\textcolor{black}{The simulated response of Chl-a concentration to temperature across the three reservoirs is presented in Fig. \ref{S1-S3}. The model predicts a unimodal relationship, with Chl-a increasing to an optimum before declining at higher temperatures. Consistent with literature values}~\citep{Gobler2020, PaerlHW2009}, \textcolor{black}{the optimal temperature range for Chl-a accumulation in our systems was approximately 20-30 \textcelsius, with peak synthesis efficiency observed between 25-28 \textcelsius. Notably, our simulations indicated sustained Chl-a growth across a broader range of 18-32 \textcelsius for reservoirs S1 and S2. Concentrations in S1 and S2 were generally higher and less sensitive to temperature variation compared to S3, where Chl-a levels showed a more pronounced response to temperature fluctuations.}
	
\begin{figure}[ht!]
		\centering
		\includegraphics[width=17cm]{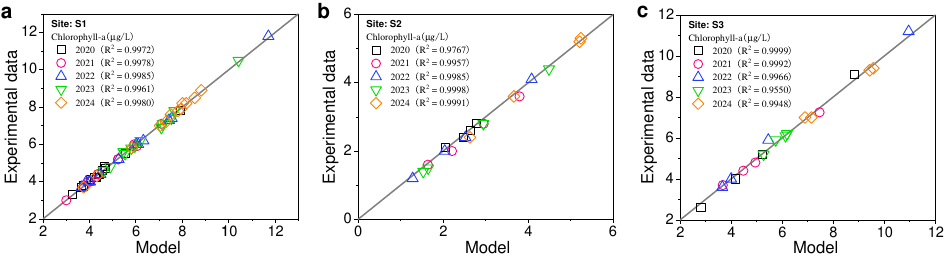}
		\caption{\label{Figdata-model}
			\textcolor{black}{Validation of model performance: comparison of simulated versus observed Chl-a concentrations across all samples from reservoirs S1, S2, and S3. The 1:1 line indicates perfect agreement. The high R\textsuperscript{2} value demonstrates the model's accuracy in simulating temperature-dependent dynamics.}
		}
	\end{figure}

	\textcolor{black}{The model's ability to capture site specific Chl-a dynamics is validated in Fig.~\ref{Figdata-model}, which compares simulated against observed concentrations across the full temperature spectrum. The high degree of agreement, supported by a coefficient of determination (R\textsuperscript{2}) close to 1, confirms that the temperature-dependent formulation in our model accurately reproduces the observed patterns. This close alignment underscores the model's utility in parsing the temperature, driven component of Chl-a variability from other interacting environmental factors.}
	
	\section{Discussion}
	\textcolor{black}{Our long-term monitoring confirms a trend of increasing eutrophication pressure in the studied reservoirs, consistent with regional observations in subtropical Asia}~\citep{Mpakairi2024,Latwal2024,HanxiaoZhang2024,Shi2024,Buta2023,	Ishikawa2022,Cairo2020}. \textcolor{black}{This widespread phenomenon is often driven by intensified anthropogenic activities and climatic shifts across the region}~\citep{Huang2022}. \textcolor{black}{The identified dominance of TN over TP as a driver of Chl-a aligns with findings from some nitrogen-rich subtropical systems but contrasts with the classical phosphorus-limitation paradigm for freshwater. This highlights the need for region-specific nutrient management strategies, as also suggested by studies in other Asian water bodies}~\citep{SaidMuhammad2024}. \textcolor{black}{For instance, many studies in a Brazilian subtropical reservoir also reported complex nutrient dynamics where phytoplankton in the lacustrine zone depended more on internal loading after riverine nutrients were consumed upstream, suggesting that the relative importance of nitrogen and phosphorus can vary spatially within a single reservoir system and over time}~\citep{Ishikawa2022,Souza2016,Cunha2016,Borges2008}. \textcolor{black}{Similar spatial complexities in nutrient-driven algal dynamics have been observed in diverse systems, from Turkish lakes to high-altitude Pakistani lakes}~\citep{CemTokatli2023, SaidMuhammad2024}.
	
	The developed dynamic model effectively integrates the key drivers. \textcolor{black}{Its performance is comparable to other recent mechanistic models applied in lakes Taihu and Chagan in China}~\citep{Liu2021, JingQian2024}. \textcolor{black}{However, our model's simplicity is also a limitation. It does not explicitly include processes such as zooplankton grazing, complex phytoplankton community shifts, or the effects of extreme precipitation events on external loading, all identified as important in other studies}~\citep{SusanneMenden-Deuer2025}. \textcolor{black}{The cascading effects of climate change on aquatic communities, including altered grazing pressure and habitat structure, represent additional complexities not captured in our current formulation}~\citep{SusanneMenden-Deuer2025}. \textcolor{black}{Furthermore, the model calibration relied on integrated surface water samples and did not incorporate vertical stratification dynamics, which can modulate nutrient and algae distributions}. \textcolor{black}{Our model, as a lumped-parameter representation, simplifies the spatially heterogeneous processes captured in remote sensing studies. Research in the Nandoni reservoir demonstrated that Chl-a concentrations can be significantly higher at reservoir edges and near inflows, a pattern driven by localized nutrient inputs and sedimentation processes that our aggregated model averages across the entire water body}~\citep{Nthunya2018}. \textcolor{black}{Similarly, the role of density currents in transporting phytoplankton from riverine to lacustrine zones, a key two-dimensional hydrodynamic process highlighted in other subtropical reservoir studies, is not explicitly resolved in our current model formulation}~\citep{Ishikawa2022}.
	
	\textcolor{black}{The anomalous peaks in TP and Chl-a observed in S2 during 2020 and 2024 underscore the significant impact of episodic events. The year 2020 saw record-breaking rainfall in the region, likely enhancing watershed runoff and non-point source pollutant transport, a phenomenon increasingly linked to climate change}~\citep{Liang2025, HulyaBoyacioglu2024}. \textcolor{black}{Such extreme events can create temporary but impactful shifts in nutrient regimes, as observed in other systems where climate-mediated changes alter nutrient export patterns}~\citep{ElisaSoana2024, ZihanZhao2025}. \textcolor{black}{The 2024 peak may be linked to recent land-use changes or specific agricultural practices upstream, highlighting the direct connection between watershed management and reservoir water quality}~\citep{ZhuoxinChen2025}. \textcolor{black}{These events highlight a key uncertainty: our model, parameterized with quarterly/monthly data, may not fully capture the system's response to such pulsed disturbances. The broader context of climate change impacts on water resources, including groundwater interactions, further complicates long-term predictions}~\citep{RicardoHirata2025}.
	
	\textcolor{black}{Despite these limitations, the model provides a valuable quantitative tool. It moves beyond statistical correlation by encapsulating key ecological mechanisms, offering a testable framework for simulating management scenarios (e.g., nutrient load reduction). The integration of long-term data with process-based modeling strengthens the causal inference regarding the roles of temperature and nutrients. This approach aligns with emerging frameworks that aim to integrate terrestrial and aquatic ecosystem responses to environmental change}~\citep{Donat-PHader2019}.
	
	This work \textcolor{black}{acknowledges several limitations and uncertainties. (1) The differing sampling frequencies between reservoirs constrain direct fine-scale temporal comparisons. (2) Although the PCA and model captured a substantial portion of the variance, the remaining unexplained variance points to the influence of unmeasured drivers, such as light availability, specific nutrient fractions (e.g., bioavailable phosphorus), or complex biotic interactions. (3) The model parameters were calibrated for the specific conditions of the studied reservoirs; therefore, their direct applicability to other systems without local validation remains an uncertainty. (4) The current modeling framework does not incorporate projections of future climate, which is a key factor for long-term predictions given the sensitivity of algal growth and nutrient cycles to warming and hydrological changes}~\citep{Gobler2020, PhilFong2025,ElisaSoana2024}. \textcolor{black}{These limitations underscore the critical need for the integrated and expanded research pathways detailed in the dedicated \textit{Future Perspectives} section. Advancing along these multifaceted lines is imperative to formulate resilient and sustainable management strategies for freshwater ecosystems facing escalating pressures}~\citep{PengQi2025}.
	\textcolor{black}{\section{Future Perspectives}}
	\textcolor{black}{This work opens several avenues for future research. (1) The differing sampling frequencies limit fine-scale comparison between reservoirs. Future monitoring should adopt harmonized, high-frequency (e.g., bi-weekly or continuous sensor-based) sampling to better resolve short-term dynamics and event responses, as emphasized in studies of long-term reservoir water quality evolution~\citep{Buta2023}. (2) The PCA and model explain a substantial portion but not all of the observed variance, indicating missing drivers. Future model iterations should incorporate additional factors such as photosynthetically active radiation (PAR), silicate concentrations, zooplankton grazing pressure, and microbial loop dynamics, which are known to mediate phytoplankton community structure~\citep{SusanneMenden-Deuer2025}. (3) The model parameters are site-specific; transferability to other subtropical reservoirs requires further validation with independent datasets from diverse systems. (4) Incorporating future climate projections (e.g., increased temperature and altered precipitation patterns from CMIP6 models) into the dynamic framework is critical for forecasting long-term ecosystem states and resilience, given the profound influence of climate on algal blooms and nutrient cycling \citep{Gobler2020, PhilFong2025, ElisaSoana2024}. (5) Expanding the model to a spatially distributed framework~\citep{Tague2004}, potentially integrated with remote sensing data for surface Chl-a validation~\citep{Nthunya2018}, would allow for the simulation of horizontal heterogeneity and identification of critical source areas within reservoirs. (6) Ultimately, linking this reservoir model with watershed-scale nutrient loading models (e.g., SWAT) would create a powerful integrated tool for evaluating the effectiveness of land-based management actions on downstream water quality, aligning with integrated catchment management approaches~\citep{ZhuoxinChen2025}.}\\
	\section{Conclusions}
	\textcolor{black}{This study integrated five years of monitoring data with a dynamic hydro-ecological model to investigate Chl-a dynamics in three subtropical reservoirs. The key findings are:} \textcolor{black}{(1) Long-term monitoring revealed a clear trend of increasing eutrophication pressure. Chl-a concentrations ranged from 1.2 to 11.8 $\mu$g/L and showed a significant upward trajectory over the study period across all reservoirs, with the most pronounced increases in Tiantangshan (S1) and Baisha River (S2) reservoirs.} \textcolor{black}{(2) Multivariate statistical analysis and dynamic modeling identified water temperature, TN, and TP as the primary co-varying drivers. Notably, TN exhibited a stronger association with Chl-a trends than TP in these specific systems, suggesting nitrogen management is a critical concern.} \textcolor{black}{ (3) The developed dynamic hydro-ecological model successfully captured the non-linear interactions among these factors. The model achieved high accuracy in simulating observed Chl-a patterns (R\textsuperscript{2} > 0.85) and identified a synergistic effect where high water temperature (> 25 \textcelsius) amplified the growth response to elevated TN.} \textcolor{black}{(4) The model provides a mechanistic, quantitative tool that moves beyond correlation. It offers a validated framework for simulating scenarios, such as evaluating the potential water quality benefits of reducing watershed nitrogen loads, thereby supporting targeted, science-informed management strategies for subtropical reservoirs.}
	
	\textcolor{black}{Overall, this study underscores the value of combining long-term observational data with process-based dynamic modeling to understand and predict complex eutrophication dynamics. It highlights the necessity of dual-nutrient (N and P) strategies and temperature-aware management in the sustainable governance of freshwater resources in rapidly developing subtropical regions under climate change.}
	
	\section*{Acknowledgments}
	This work was supported by Jiangsu Provincial 'Double Innovation' Doctoral Talent Fund (No. JSSCBS0620) and Jiangsu University of Science and Technology Young Teachers Research Initiation Fund (No. 1202932307). 
	
	\section*{Author contributions}
	\textbf{Haizhao Guan:} Writing – review \& editing, Writing – original draft, Visualization, Methodology, Investigation, Formal analysis, Data curation, Conceptualization. \textbf{Yiyuan Niu:} Writing – review \& editing, Writing – original draft, Visualization, Methodology, Investigation, Formal analysis. \textbf{Chuanjin Zu:} Writing – review \& editing, Investigation, Funding acquisition. \textbf{Ju Kang:} Writing – review \& editing, Writing – original draft, Visualization, Validation, Software, Methodology, Investigation, Formal analysis, Data curation, Conceptualization.
	
	\section*{Competing interests}
	The authors declare that they have no conflict of interest.
	
	\section*{Code and data availability} 
	Data and code will be made available on request.
	%\bibliographystyle{customnat}
	%\bibliographystyle{plainnat}%plainnat
	%\bibliography{reference}

\end{document}